\documentclass[journal]{IEEEtran}

\usepackage{epsf}
\usepackage{graphicx}
\usepackage{amssymb}
\usepackage{subfigure}
\usepackage{array}

\begin{document}

\title{Modal  control in semiconductor optical waveguides with
uniaxially patterned layers}
\author{Arsen~V.~Subashiev, and
Serge~Luryi,~\IEEEmembership{Fellow,~IEEE}
\thanks{Based on manuscript submitted May 3, 2005 to IEEE J. of Lightwave Technology.
       This work was supported by the NY State Center
       for Advanced Sensor Technology (Sensor CAT) at Stony Brook.}%
\thanks{Arsen~V.~Subashiev is with the
Department of Electrical and Computer Engineering, State University
of New York at Stony Brook, Stony Brook, NY, 11794-2350, on leave
from the State Polytechnic University, St. Petersburg, Russia, 195251
(e-mail:subashiev@ece.sunysb.edu).}
\thanks{Serge~Luryi is with the
Department of Electrical and Computer Engineering, State University
of New York at Stony Brook, Stony Brook, NY,
11794-2350 (e-mail:Serge.Luryi@stonybrook.edu).} \markboth{Journal of
\LaTeX\ Class Files,~Vol.~1, No.~11,~November~2005}} \maketitle
\begin{abstract}
Uniaxially patterned dielectric layers have an optical anisotropy
that can be externally controlled.
We study the effects of patterning the cladding
or the core layer of a 3-layer optical waveguide
on the polarization properties of propagating
radiation. Particular attention is paid to the case
when the core material is a semiconductor with optical gain.
We discuss a number of devices based on incorporating an uniaxially
patterned layer in the structure design, such as a
polarization-insensitive amplifier, a polarizer, an
optically-controlled polarization switch, and an optically
controlled modal coupler.
\end{abstract}


\section{Introduction}

\PARstart{S}{tructures} with cylindrical air pores forming a 2-d
periodic lattice in a semiconductor material are actively studied
for photonic bandgap applications, \cite{Yablonov,Painter,Janopp}
such as spontaneous emission control and light confinement in
micro-cavities. These studies stimulated numerous computations of
the photonic crystal (PC) band spectra, based on the plane wave
expansion of the electromagnetic field \cite{Maradu,Joannop}. Such
calculations showed that in the long wavelength limit the spectrum
of electromagnetic waves can be well described in the effective
media approximation with an effective dielectric constant
corresponding to the results of Maxwell Garnett theory
\cite{MG,Nicoro,Sarychev}. Optical properties of the composite
structures patterned with cylindrical holes, for the wavelentgh
$\lambda$ exceeding the interhole spacing $a$, i.e., for $\lambda
\gg a$,  are described in terms of the filling factor $f$ alone
(i.e. the fraction of the total volume occupied by the pores), and
do not depend on the long-range order of the holes or their
diameter. The effect of disorder is only a weak Rayleigh-like
scattering. The effective media approach remains valid for very
large contrast ratios between the semiconductor and the pore
permittivities \cite{Driel,Halevi,Halevi2} and for arbitrary
propagation directions of the electromagnetic waves. Direct
comparison of the calculation results based on 3-d and 2-d modeling
shows that the same approach can be used to describe the waveguiding
properties of multi-layered structures that include patterned
layers. Moreover, studies of PC-like structures with a small
disorder showed that the Maxwell Garnett approach remains valid even
when the requirement $\lambda \gg a$ is relaxed to $\lambda > a$, so
long as the optical frequency is below the photonic bandgap and
light scattering remains negligible (similar situation prevails in
electronic spectra engineering with quantum well and superlattice
heterostructures).

In this paper we explore variable anisotropic optical properties of
uniaxially patterned (UAP) layers and find that they can be useful
in the design of numerous optical devices that are sensitive to the
shape and polarization of the optical mode, such as polarizers,
lasers, amplifiers and modulators. The UAP anisotropy is not
accompanied by any additional optical loss and therefore can be used
effectively for the modal control of optical emitters and
amplifiers.

Polarization  sensitivity is an important factor in  semiconductor
lasers and amplifiers. It depends on the modal gain which in turn
depends on both the material gain anisotropy and the mode
confinement factor \cite{Adams}. The traditional 3-layer waveguide
design of semiconductor amplifiers with isotropic constituents leads
to a better confinement of the TE  mode and a larger modal gain for
this mode compared to the TM mode \cite{Visser,Huang}. To obtain a
polarization-insensitive amplifier one had to use highly anisotropic
active layers with the material gain that favors TM polarization.
Adoption of UAP media for the waveguide layers, gives an additional
possibility to compensate for the difference in the TE-TM
confinement, inherent to the isotropic situation.

Possible applications of the waveguide structures with a UAP layer
extend to the territory already tested experimentally for PC layers,
such as structures with a periodically patterned cladding layer,
e.g. \cite{Painter} and periodically patterned active layer, e.g.
\cite{Xu,Wu}. For the UAP structures we consider, the pattern
long-range order is of no consequence. The relative value of modal
propagation constants can be altered by varying the thickness of the
core region or by varying the fill-factor of the patterned layer.
The propagation constants can be further fine-tuned by changing the
optical contrast between the waveguide constituents with an applied
field or optical pumping. Tuning effects are enhanced in structures
that are particularly sensitive to the anisotropy of each layer,
such as asymmetric waveguides with a thin core layer.

\section{Dielectric function of uniaxially patterned layers and the
waveguide modes}

Anticipating a broad scope of possible applications we consider a
3-layer waveguide in which all three layers, the top cladding (c),
the core (f), and the substrate or bottom cladding (s), may be
uniaxially patterned. We assume a thin core layer (of thickness $d$)
that can support only the lowest propagation modes. We examine the
case when the optical axis $C$ of the patterned layers is
perpendicular to the waveguide plane. We denote by
$\epsilon_{\|,{c,f,s}} $ and $\epsilon_{\perp,{c,f,s}}$ the
permittivities of the $c$s, $f$, or $s$ layers for two directions of
the electric field, parallel ($\parallel$) and perpendicular
($\perp$) to the optical axis, respectively. The permittivity of the
inhomogeneous medium in a long-wavelength limit is obtained in the
Maxwell Garnett approximation.

For $s$-polarization (${\bf E} \parallel C$), the permittivity of a
2-d array of infinitely long cylinders is obtained by direct
averaging, viz.
\begin{equation}
\epsilon_{ \|} =  \epsilon_{out} + (\epsilon_{in} - \epsilon_{out})
f , \label{eps_par}
\end{equation}
where the permittivity $\epsilon_{in}$ is inside and
$\epsilon_{out}$ outside the cylinder.

For $p$-polarization (${\bf E} \perp C$) the permittivity is given
by

\begin{equation}
\epsilon_{ \perp} =  \epsilon_{out} {{ (\epsilon_{in} +
\epsilon_{out}) +( \epsilon_{in}- \epsilon_{out})f} \over {
(\epsilon_{in} + \epsilon_{out})-( \epsilon_{in}- \epsilon_{out})f}}
. \label{eps_perp}
\end{equation}
\begin{figure}
\leavevmode \centering{
\includegraphics[width=2.7in]{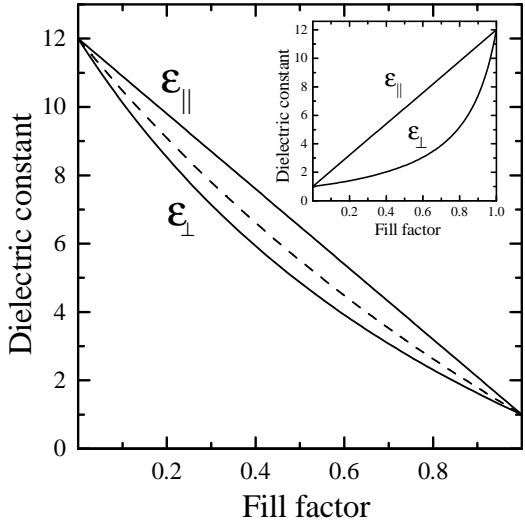}}
\caption{Average permittivities $\epsilon_{\|}$ and
$\epsilon_{\bot}$ of a silicon ($\epsilon_{out}=12$) layer
uniaxially patterned with cylindrical air pores ($\epsilon_{in}=1$)
for ${\bf E} \parallel C$ and ${\bf E} \bot C $, respectively. The
dotted line shows the average refractive index (squared). The inset
shows the $\epsilon_{\|}$ and $\epsilon_{\bot}$ for a mirror
structure of dielectric cylinders, ($\epsilon_{in} \rightleftarrows
\epsilon_{out}$).}
\end{figure}

The dependence of $\epsilon_{\|}$ and $\epsilon_{\perp}$ on the
filling factor  is shown in Fig. 1 for the case of cylinder pores in
a dielectric medium, with $\epsilon_{in}=1$ and $
\epsilon_{out}=12$. The anisotropy of refractive index is evidently
not small, e.g., for $f = 0.3$, one has $(n_{\|} - n_{\perp})/<n>
=0.1$). Also shown is the value of permittivity that would
correspond to an average refractive index, i.e. $<n> = n_{in} f +
n_{out}(1-f)$ which is sometimes used, see e.g. \cite{Charlton},
without a reasonable justification \cite{average}.

The inset to Fig. 1 shows the permittivities $\epsilon_{\|}$ and
$\epsilon_{\perp}$ for a "mirror" array of cylindrical rods in air.
This geometry offers a substantially higher optical anisotropy.
Mathematically, it is described by Eqs. (\ref{eps_par}) and
(\ref{eps_perp}) with the replacement $\epsilon_{out}
\rightleftarrows \epsilon_{in}$.

Note that Eq. (\ref{eps_perp}) fails for thin UAP layers, when the
hight of the cylinders becomes comparable to their diameter. For
this case the effective media approach remains valid, but Eq.
(\ref{eps_perp}) must be modified to allow for depolarization
factors of the finite-height cylinders.

Below, we discuss properties of a 3-layer waveguide. We shall employ
the usual approach \cite{Adams} developed for isotropic waveguide
constituents. The guided modes supported by the structure will be
calculated using the values $n_{\|}$ and $n_{\perp}$ as
polarization-dependent refractive indices of the patterned layers.
Exemplary profiles of the dielectric function are shown in Fig. 2
for a waveguide with patterned cladding layer \emph{(a)}, an
asymmetric waveguide with a UAP core layer, \emph{(b)}, and a
hypothetical structure with both cladding and
core layers patterned \emph{(c)}. %

\subsection{TE mode}
\label{TE}
\begin{figure}[tb]
\leavevmode \centering{
\includegraphics[width=3.3in]{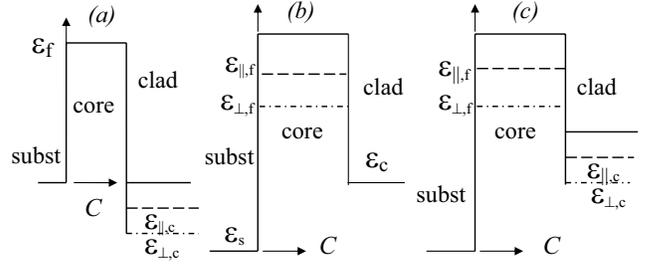}}
\caption{Profile of the dielectric function in the 3-layers
waveguide with PL layers: patterned cladding layer (a), patterned
core layer (b) and patterned core and cladding layers (c). }
\end{figure}

In the TE mode the electric field is perpendicular to the pore axes,
so that {\em ordinary} waves propagate in all 3 layers. Therefore,
in order to find the modal index, we can use the eigenvalue equation
for the propagation wave vector in an isotropic layered structure
\cite{Adams} %
\begin{equation}
k_{o,f} d = \tan^{-1}(\alpha^{c}/k_{o,f})+\tan^{-1}
(\alpha^{s}/k_{o,f}) \label{TE-Deq}
\end{equation}

\noindent in which the substitution $ \epsilon_{c,f,s} \rightarrow
\epsilon_{\perp,{c,f,s}}$  should be made, viz., where $k_{o,f} =
\sqrt{ \epsilon_{\perp,f}k_0^2 - Q^2}$, $\alpha^{c} =
\sqrt{Q^2-\epsilon_{\perp,c}k_0^2 }$, $\alpha^{s}=
\sqrt{Q^2-\epsilon_{\perp,s}k_0^2 }$, $k_0  =  \omega/c$, and $Q$ is
the propagation wave vector, which defines the mode effective index,
$n_{eff} = Q/k_0$.

From Eq. (\ref{TE-Deq}) the cutoff thickness $d_{c,TE}$ of the
active layer for the case $\epsilon_{\perp,s}>\epsilon_{\perp,c}$ is
given by %
\begin{equation}
d_{c,TE} = \left[ k_0 \sqrt{(
\epsilon_{\perp,f}-\epsilon_{\perp,s})}
\right]^{-1}\tan^{-1}(\sqrt{a_{TE}}) , \label{cutoff_te}
\end{equation}

\noindent where $a_{TE}$ is an asymmetry parameter of the form %
\begin{equation}
a_{TE} = {{\epsilon_{\perp,s}-\epsilon_{\perp,c}} \over {
\epsilon_{\perp,f}-\epsilon_{\perp,s}}} \label{asym_te}
\end{equation}

\noindent The case $\epsilon_{\perp,s} < \epsilon_{\perp,c}$ is
described by replacing  $\epsilon_{\perp,s} \rightleftarrows
\epsilon_{\perp,c}$.

In the limit of a very thin active layer, $k^f
d \ll 1$, equation (\ref{TE-Deq}) yields %
\begin{equation}
n_{eff,TE} = \sqrt{\epsilon_{\perp,s}}+
{\delta^2_{TE}\over{2\sqrt{\epsilon_{\perp,s}}}} ,
\end{equation}
where
\begin{equation}
\delta_{TE}= {{k_0^2d^2(\epsilon_{\perp,f}-\epsilon_{\perp,s})^2 -
(\epsilon_{\perp,s}-\epsilon_{\perp,c}) }\over {2 k_0d
(\epsilon_{\perp,f}-\epsilon_{\perp,s})}}~ \ll 1 . \label{ind_TE}
\end{equation}

For an amplifying structure with active core, the gain factor for
the TE mode $g_{TE}$ can be calculated as follows
\cite{Visser}: %
\begin{equation}
g_{TE} =  - { k_0 \over n_{eff,TE}}%
{{ \int_0^d dx \epsilon''_{f} |E_y|^2 }\over {\int_{-
\infty}^{\infty} dx |E_y|^2}} , \label{gain_TE}
\end{equation}

\noindent where $\epsilon''_{f}$ is an imaginary part of the active
layer permittivity, $x$ is taken along $C$ and $y$ perpendicular to
the wave propagation direction. For a UAP core, the integration in
Eq. (\ref{gain_TE}) includes taking the average over the layer
plane. The modal gain (\ref{gain_TE}) can be written as a product of
the material gain $g_f = k_0 \epsilon''_{f}
/\sqrt{\epsilon_{f,out}}$ and the optical confinement factor,
$\Gamma_{TE}$, which can be calculated explicitly in the dipole
approximation for the field distribution in the patterned layer. For
a structure with a thin patterned core layer, $\Gamma_{TE}$ equals

\begin{equation}
\Gamma_{TE} = s{\sqrt{\epsilon_{f,out} \over {
\epsilon_{s}}}}{{2\sqrt{\epsilon_{s}-\epsilon_{c}+
\delta^2_{TE}}}\over
{\sqrt{\epsilon_{s}-\epsilon_{c}+\delta^2_{TE}}+
\delta_{TE}}}\delta_{TE} k_0 d , \label{Conf_TE}
\end{equation}

\noindent where $s$ is the local field factor $s
=(1-f+B^2f)/(1+Bf)^2$, and
$B=(\epsilon_{f,out}-\epsilon_{f,in})/(\epsilon_{f,out}+
\epsilon_{f,in}) $. For a symmetric waveguide, Eq. (\ref{Conf_TE})
reduces to the well-known result \cite{Dumke}. For an asymmetric
waveguide, $\delta_{TE}$ rapidly decreases with the difference of
the indices of the cladding and substrate layers, as follows from
Eq. (\ref{ind_TE}). This leads to a high sensitivity of the optical
confinement to both the asymmetry of the waveguide and the layer
anisotropy.

\subsection{TM mode}

For the TM mode the electric field has two components, one
perpendicular and the other parallel to the $C$ axis. The
propagating waves are {\em
extraordinary} in the UAP layers. The eigenvalue equation is of the form %
\begin{equation}
k_{e,f} d = \tan^{-1}\left({{\epsilon_{\perp,f}\beta^{c}} \over
{\epsilon_{\perp,c}k_{e,f}}}\right)+ \tan^{-1}
\left({{\epsilon_{\perp,f}\beta^{s}}\over {
\epsilon_{\perp,s}k_{e,f}}}\right) , \label{TM-deq}
\end{equation}

\noindent where $$k_{e,f} = \sqrt{ \epsilon_{\perp,f}k_0^2 -
(\epsilon_{\perp,f}/\epsilon_{\|,f}) Q^2},$$  $$\beta^{c} =
\sqrt{(\epsilon_{\perp,f}/\epsilon_{\|,f}) Q^2-
\epsilon_{\perp,c}k_0^2},$$ and $$\beta^{s}=
\sqrt{(\epsilon_{\perp,f}/\epsilon_{\|,f})Q^2-
\epsilon_{\perp,s}k_0^2 }.$$ \noindent The cutoff thickness
$d_{c,TM}$, for a structure with $\epsilon_{\|,s}>\epsilon_{\|,c}$
is given by %
\begin{equation}
d_{c,TM} = \left[ k_0 \sqrt{ (\epsilon_{\perp,f}/ \epsilon_{\|,f})~(
\epsilon_{\|,f}  -\epsilon_{\|,s})}\right]^{-1}\tan^{-1}
(\sqrt{a_{TM}}) , \label{cutoff_tm}
\end{equation}

\noindent where the asymmetry parameter $a_{TM}$ has the form %
\begin{equation}
a_{TM} = {{\epsilon_{\perp,f}\epsilon_{\|,f}~(\epsilon_{\|,s}-
\epsilon_{\|,c})}\over {\epsilon_{\perp,c}\epsilon_{\|,c}~
(\epsilon_{\|,f}-\epsilon_{\|,s})}} \label{asym_tm}
\end{equation}

\noindent The case $\epsilon_{\|,s} < \epsilon_{\|,c}$ is described
by replacing $\epsilon_{\|,s} \rightleftarrows \epsilon_{\|,c}$ and
$\epsilon_{\perp,s} \rightleftarrows \epsilon_{\perp,c}$.

\noindent For a very thin active layer, $k_{x,o}^f d \ll 1$, Eq.
(\ref{TM-deq}) yields %
\begin{equation}
n_{eff,TM} = \sqrt{ \epsilon_{\|,s}}+ {\delta_{TM}^2 \over {2 \sqrt{
\epsilon_{\|,s}}}},
\end{equation}

where

\begin{equation}
\delta_{TM} = {{k_0^2d^2r_1^2(\epsilon_{\|,f}-\epsilon_{\|,s})^2 -
r_2^2(\epsilon_{\|,s}-\epsilon_{\|,c}) }\over {2 k_0d r_1
(\epsilon_{\|,f}-\epsilon_{\|,s})}},~~~~\delta_{TM}\ll1
\label{ind_TM}
\end{equation}
\noindent with
$r_1=\sqrt{\epsilon_{\|,s}\epsilon_{\perp,s}}/\epsilon_{\|,f}$ , and
$r_2=\sqrt{\epsilon_{\|,s}\epsilon_{\perp,s}/\epsilon_{\|,c}
\epsilon_{\perp,c}}$.

In calculations of the  modal gain $g_{TM}$  and the optical
confinement $\Gamma_{TM}$ for the TM mode one must correctly
evaluate  the energy flux in and outside the active layer
\cite{Visser}. We write down the optical confinement factor for a
structure with a thin patterned core layer,  %
\begin{equation}
\Gamma_{TM} =(1-f) { {\epsilon_s \sqrt{\epsilon_s
\epsilon_{f,out}}}\over \epsilon^2_{\|,f}}
{{2\sqrt{\epsilon_{s}-\epsilon_{c}+ \delta^2_{TM} }}\over
{\sqrt{\epsilon_{s}-\epsilon_{c}+ \delta^2_{TM}}+ \delta_{TM} }}
\delta_{TM}k_0d \label{Conf_TM}
\end{equation}

\noindent As follows from Eqs. (\ref{ind_TM},\ref{Conf_TM}), due to
a small multiplier $r_1^2$ in the numerator of (\ref{ind_TM}), both
$\delta_{TM}$ and hence $\Gamma_{TM} $ are even more sensitive to
the asymmetry and anisotropy than the analogous parameters for the
TE-mode. When the active layer is thin, $k^f_{x}d\ll 1$, the
confinement for the TM mode is smaller by a factor of
$\approx(\epsilon_{s}/\epsilon_{f})^3$ than that for the TE mode.
For thicker layers, $k^f_x d \sim 1$, this ratio  reduces to
$(\epsilon_{s} /\epsilon_{f})^2$, see e.g. \cite{Visser,Huang}.

The mode competition in laser structures is also affected by the
difference in the reflection coefficients for the competing modes.
For a cleaved stripe structure the modal reflection coefficients
$R_m$
are given by \cite{Reflec} %
\begin{equation}
R_m = {(n_{eff,m} - 1 )^2\over (n_{eff,m} + 1 )^2},~~~~~~m={\rm
TE,TM}, \label{CleavedStripe_R}
\end{equation}

Thus, in the effective index approach this ratio is a function of
$n_{eff,TE}$ and $n_{eff,TM}$ and, therefore, is also affected by
the fill factor of the UAP layers of the waveguide.

\section{Possible applications of UAP structures}

We  have shown that the cutoff thicknesses and modal propagation
constants in waveguides with a thin core layer are sensitive to the
permittivities of the layers and their patterning. Small variations
of the propagation constants result in substantial changes of the
confinement factors modal ratio. This modal control can be employed
in optical devices, such as polarizers and mode-insensitive
amplifiers. It is important to realize that the control can be
effected rapidly and in real time. For example, optical pumping of
the UAP layer within the absorption band of one of its constituent
materials will change the optical contrast of the uniaxial pattern
and thus modify both the refractive index of the UAP layer and the
modal indices of the waveguide. Thus, we can have an ultra-fast
switch of the modal response in an anisotropy-based cut-off device.
Other possible applications are mode-dependent leaky waveguides and
directional couplers. With an additional high-index layer, adjacent
to one of the cladding layers, the coupling of waveguide modes to
this layer will have a strong dependence on the matching of modal
propagation constants.  %

\subsection{Mode tuning and polarization-insensitive amplifier}

\begin{figure*}[tb]
\centering{\subfigure{\includegraphics[width=2.6in]{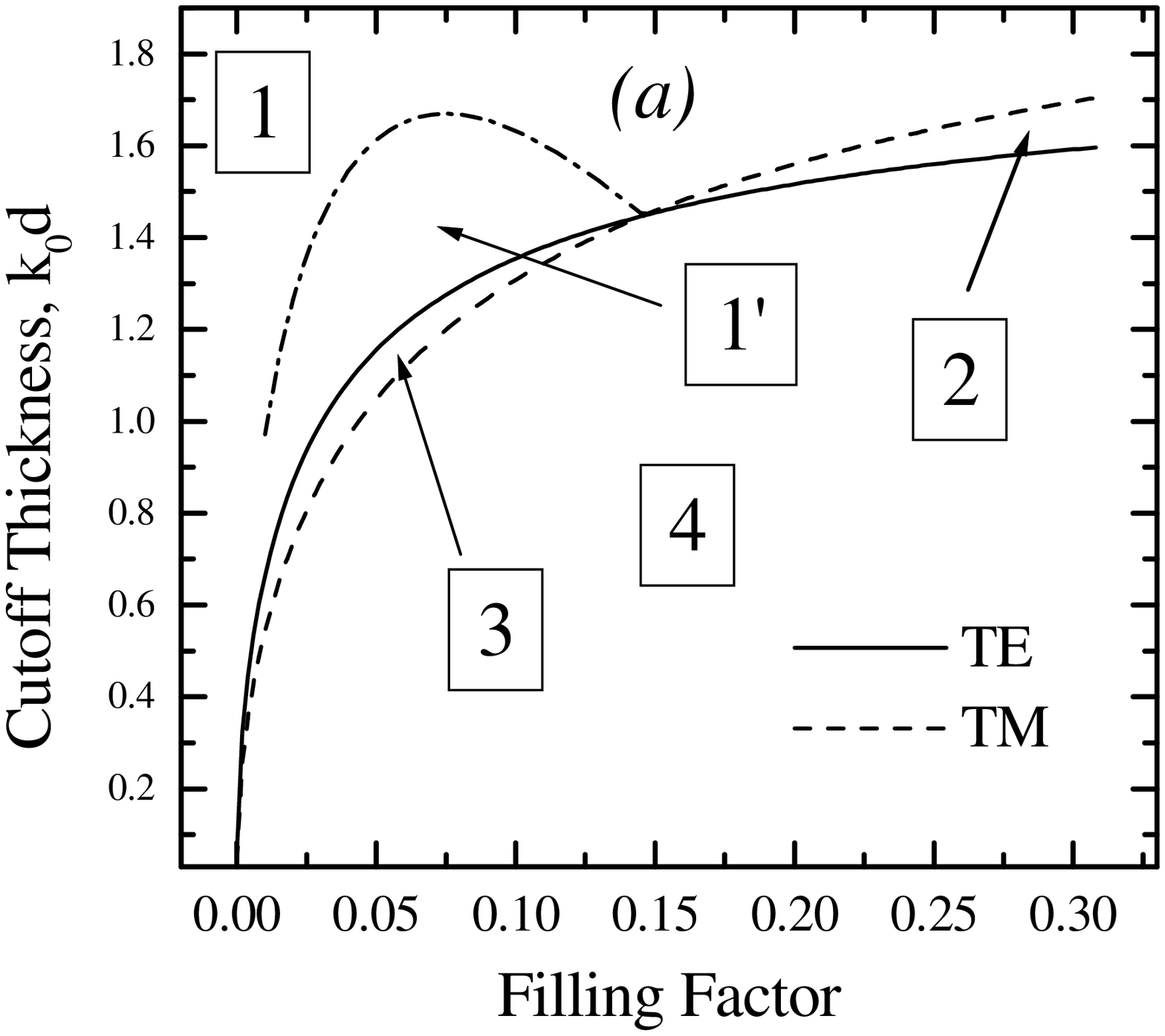}}
~~~~~ \subfigure{\includegraphics[width=2.9in]{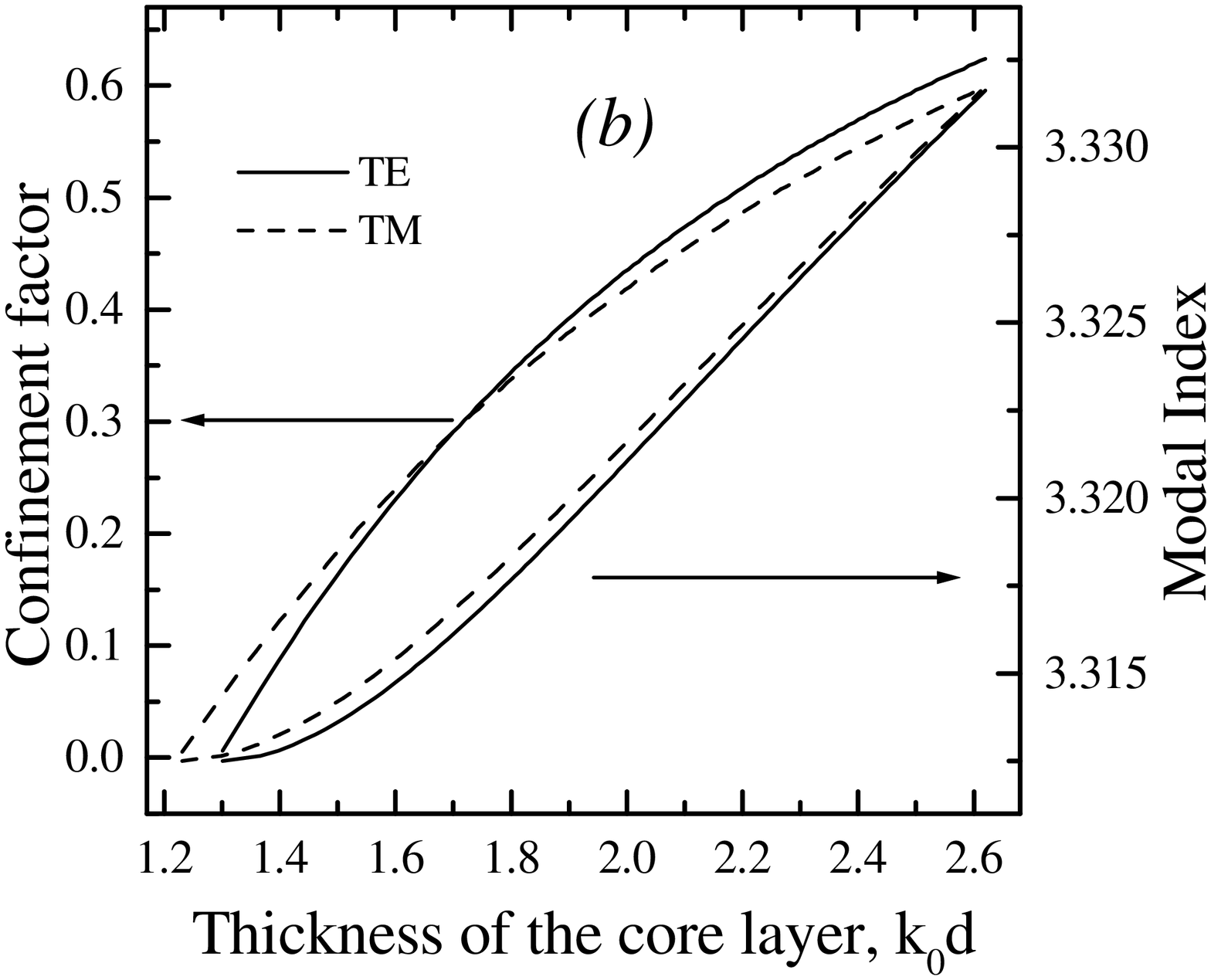} }}
\caption{(a) Cutoff thicknesses for TE and TM modes as functions of
the fill-factor in the UAP cladding layer for
Al$_{0.17}$Ga$_{0.83}$As/GaAs/Al$_{0.17}$Ga$_{0.83}$As structure.
Four distinct mode confinement regions are identified: (1) both
modes are confined, (2) only TE mode is confined, (3) only TM mode
is confined, and (4) both modes are not confined. The dash-dotted
line corresponds to the locus of points on the graph, where the
confinement of both modes is the same. It delineates a region
($1^{\prime}$) within (1) where $\Gamma_{TM}>\Gamma_{TE}$. (b)
Variation of the optical confinement factors and the modal indices
of the two modes with the active layer thickness $d$ in a UAP
structure as in (a) for a fixed filling factor $f$=0.08. Equal modal
gain at $\lambda=$860 nm is achieved for $d$ = 182 nm.}
\end{figure*}

To clarify the effects of a UAP layer on the waveguide modal
properties, we consider the cutoff thickness of a symmetric 3-layer
waveguide, in which one of the cladding layers is patterned, cf. the
index profile of Fig. 1(a). Exemplary material compositions are
taken for a GaAlAs heterostructure, specifically, GaAs core and
Al$_{x}$Ga$_{1-x}$ with $x$=0.17 for both the cladding and the
substrate layers. We assume the UAP structure in the cladding layer
(cf. the structure of Ref. \cite{Painter}). The alloy refractive
index is taken in the form $n(x)=3.4-0.53x+0.09x^2$, \cite{Ioffe}.

Variation of the cutoff thicknesses $d_{c,TE}$ and $d_{c,TM}$ with
the fill factor is shown in Fig. 3(a) in units of $1/k_0$.  Both
modes are confined in region 1, and neither mode is supported in
region 4. Region 2 supports only the lowest TE mode and region 3
only the lowest TM mode. For a fill-factor $f \le 0.141$ we see that
$d_{c,TE} < d_{c,TM}$ and we can have a waveguide which supports
only the lowest TM wave. For $\lambda =$ 0.86 $\mu$m and $f=0.08$,
the interval where this is the case is  167 nm $\le d \le$ 177 nm.
Similarly, for  $f \ge 0.141$ there is an interval of layer
thicknesses in which only TE mode is confined. The reversal of modal
confinement is due to a rapid decrease with $f$ of the cladding
layer indices for both polarizations. This leads to a better
confinement of the TE mode at large $f$, since in a strongly
asymmetric waveguide, the anisotropy is of minor importance.

The fact that $d_{c,TE} < d_{c,TM}$ in a certain range of
fill-factors indicates that there is a {\it region } of core
thicknesses {\it in the same range} where both modes are supported
but the TM mode has a tighter optical confinement. This region,
designated as $1^{\prime}$, is delineated in Fig. 3(a) by the
dash-dotted line. In the vicinity of the dash-dotted line there is
another line where gain is mode-insensitive (precise position of
this line depends on other factors, such as anisotropy of the
material gain and modal dependence of the feedback). This enables us
to design a mode-insensitive amplifier without relying on those
other factors.

It should be noted that in a waveguide with active (amplifying or
absorbing) layers, the waveguiding itself is influenced by the
gain/damping effects. For structures with a thin core layer, $k_0 d
\ll 1$, these effects, however, are smaller than the index-guiding
effects by a factor of $(k_0 d)^2$, see Appendix \ref{Appendix1},
and they can be safely neglected.

Figure 3(b) shows the variation of optical confinement factors and
effective modal indices as functions of the active layer thickness
for an exemplary fill-factor $f = 0.08$. Equal confinement is
obtained at $d = 1.62 \times k_0^{-1}$= 221.6 nm (for $\lambda_0=
860$ nm). Since at this thickness $n_{eff,TE} < n_{eff,TM}$, the
design of a mode-insensitive amplifier should also take into account
the different modal reflection coefficients, cf., e.g.,
(\ref{CleavedStripe_R}).

Pore spacings $\le 100$ nm and pore diameters $\le 30$ nm
\cite{Xu,Wu} are demanding but achievable with focused ion beam
patterning. Parameters of the structure discussed above are
adequately addressed with an approximately triangular lattice of
pores of diameter 40 nm and pitch $a$ = 134 nm. For such a lattice,
the pitch remains comfortably shorter ($\approx 2$ times) than the
wavelength in the media. Requirements to the structure parameters
are less demanding in the infrared region.

Equal modal confinement can also be obtained in the conceptually
simpler (though probably less practical) case when the UAP layer is
the core of a symmetric waveguide. This case can be easily analyzed
in a similar fashion. We find that with a thin active layer the TM
mode can be made competitive if one uses a waveguide with a
relatively small initial index contrast, which makes it more
sensitive to the core layer anisotropy. The desired low contrast is
obtained by an appropriate choice of the fill-factor of the
patterned core layer and the cladding layers composition. %

\subsection{Cutoff polarizer}

In waveguides based on III-V heterostructures, the index contrast
between the core and the cladding layers is weak. Because of this,
the modal competition takes place at small $f$ and for a thin core.
The region of competition can be made substantially larger in
asymmetric waveguides with properly chosen compositions in the
substrate and cladding layers. For the
Al$_x$Ga$_{1-x}$As/GaAs/Al$_y$Ga$_{1-y}$As waveguide structure, one
should take a smaller Al concentration $x$ in the UAP (cladding)
layer than the Al concentration $y$ in the substrate. One can then
find the fill-factor values $f_{TE},f_{TM}$, for which,
respectively, $\epsilon_{c,\perp}=\epsilon_{s}$ and
$\epsilon_{c,\|}=\epsilon_{s}$, i.e. the waveguide becomes
symmetrical for one of the waves. Figure 4 shows variation of the
cut-off thicknesses for a waveguide structure with $x=0.2$ and
$y=0.7$. In the vicinity of $f=0.085$ only the TE mode is confined
in the interval of $0 \ge d \ge d_{c,TM}$ and for $f=0.145$ only the
TM mode is confined in the interval of $0 \ge d \ge d_{c,TE}$. Thus,
the waveguide with a judiciously chosen fill-factor and active layer
thickness can be used as a cutoff-based polarizer. Moreover, region
($1^{\prime}$) can be extended to higher values of $f$ (which would
make the structure easier to make) by using a structure with both
core and cladding layers patterned. %

\begin{figure}[h]
\leavevmode \centering{
\includegraphics[width=2.7in]{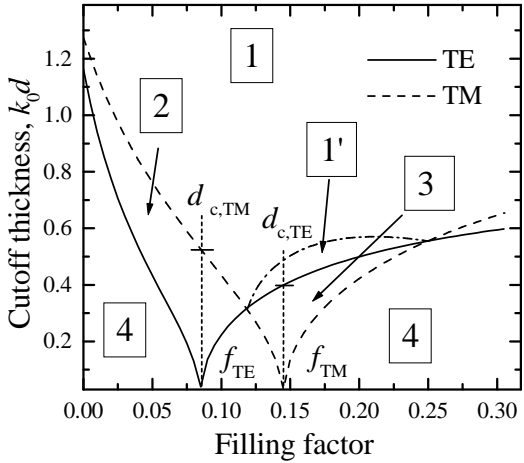}}
\caption{Cutoff thicknesses for TE and TM modes in an asymmetric
structure (Al$_x$ Ga$_{1-x}$ As / GaAs / Al$_y$ Ga$_{1-y}$As with
$x$ = 0.2, $y$ = 0.7) as functions of the filling factor of the
patterned Al$_x$ Ga$_{1-x}$ As cladding. Vertical lines indicate
when the waveguide becomes symmetric for one of the modes. Regions
are designated as in Fig. 3.}
\end{figure}

\begin{figure*}[tb]
\leavevmode \centering{
\subfigure{\includegraphics[width=2.56in]{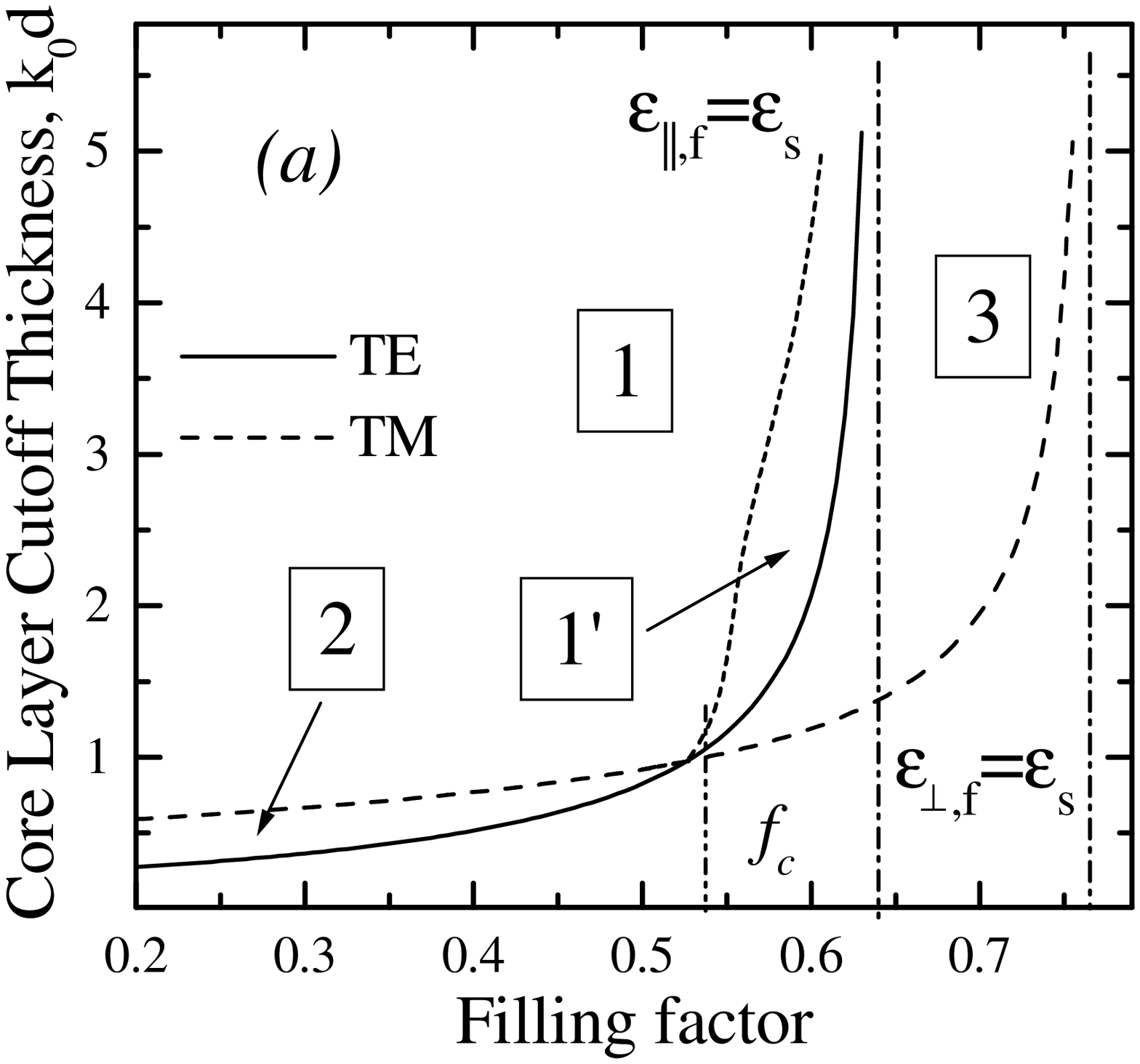}}~~~~~
\subfigure{\includegraphics[width=2.9in]{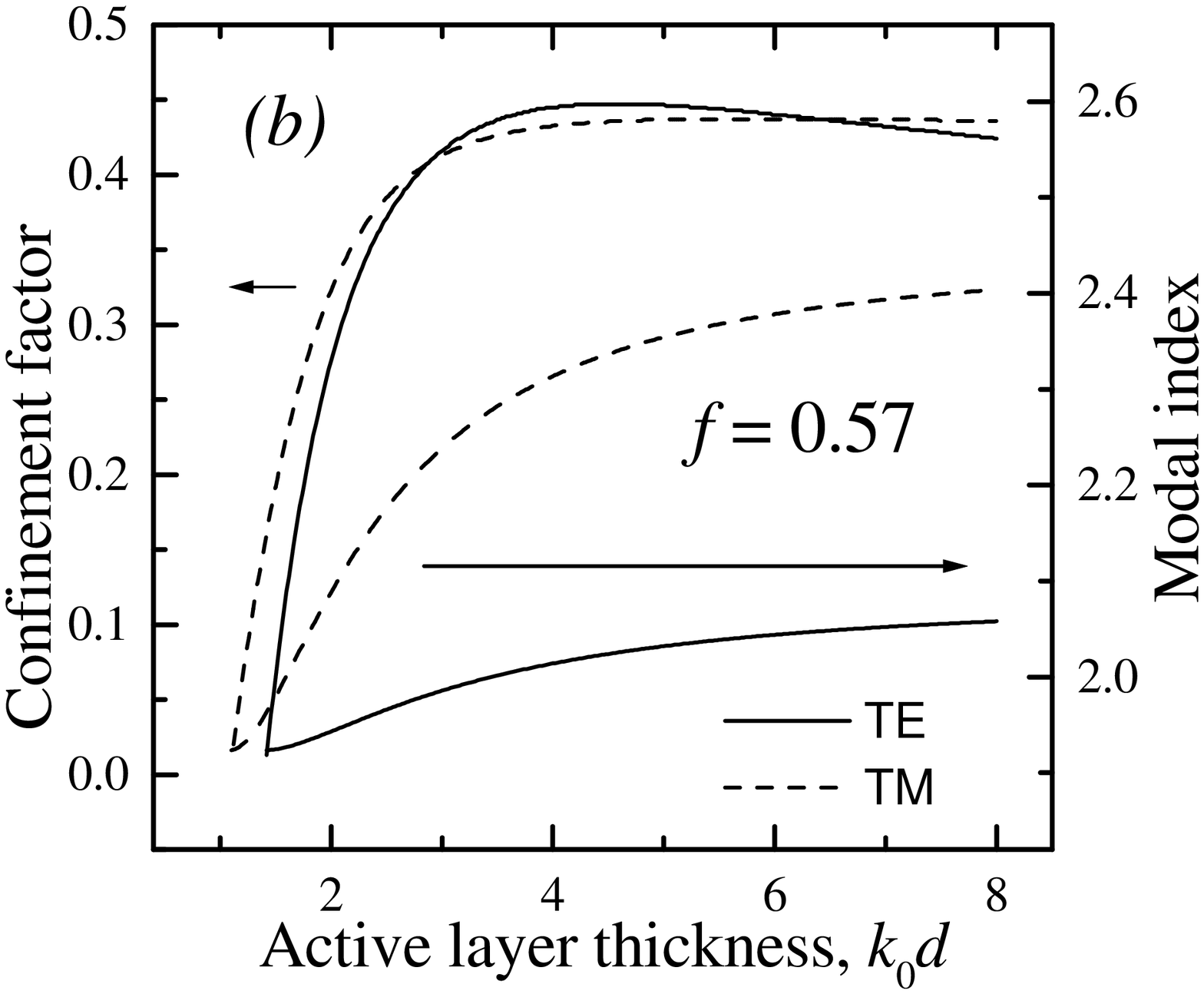}}}
\caption{Cutoff thicknesses and confinement regions for TE and TM
modes as a function of fill-factor of the UAP core layer for
asymmetric Si/SiN waveguide with air as a cladding layer (a), the
regions being noted as in Fig. 3(a); modal indexes  and confinement
factors for TE and TM modes as a function of core layer thickness
for a structure with f=0.45 (b). }
\end{figure*}

\subsection{Dominant lasing mode in highly asymmetric structures}

As an example of highly asymmetric waveguide, we consider a
structure with air for the cladding layer and a UAP semiconductor
core layer (Si, $\epsilon_{out}=12$) on a dielectric substrate (SiN,
$\epsilon_{s}=3.7$). The waveguide profile is illustrated in Fig.
2(b).

First, we calculate the cutoff thicknesses for TE and TM modes,
Their dependencies on the fill-factor of the UAP layer are shown in
Fig. 5(a). In the range $f \ge f_c=$ 0.53 the TM mode has a smaller
cutoff thickness and there is a wide range of thicknesses (region 3)
where the waveguide will support only the lowest TM mode. We see
that in strongly asymmetric waveguide structures, at large $f$, the
TM mode has better confinement and larger modal index. This results
from the faster growth with $f$ of the asymmetry factor $a_{TE}$,
compared to $a_{TM}$, see Eqs. (\ref{asym_te}) and (\ref{asym_tm}).
For values of the fill-factor near $f_c$ both modes have a similar
confinement factor in a broad range of $d$, as illustrated in Fig. 5
(b). Note that the values of the confinement factors are generally
reduced due to porosity of the core layer.

The examined asymmetric waveguide is similar to that used by
Cloutier and Xu \cite{Xu} who observed a predominantly TM-polarized
laser-like emission from a UAP Si-on-insulator layer. The main
difference from Fig. 5 is that a lower-index SiO$_2$ was used as the
bottom cladding. It would be tempting to seek an explanation for the
observed TM polarization in terms of the UAP properties of the
waveguide used. However, the structure parameters in Ref. \cite{Xu}
correspond to $f = 0.18 < f_c$ and for the stated core-layer
thickness fall within region 2 of Fig. 5(a). Not only is the TE mode
"better" confined, but the TM mode is {\it not confined at all} at
the operating point. Therefore, the observed TM polarization of the
generated light in the experiment \cite{Xu} poses a serious problem,
see Appendix \ref{Appendix}.

\subsection{Polarization switch}

Under high illumination the photo-induced concentration of free
electrons in the core and/or cladding layer(s) can be large enough
for a substantial change of the permittivity and thus effect a
change of the modal confinement in a UAP waveguide. Using materials
with a short carrier lifetime, both the rise time and the recovery
time can be very short, thus providing an ultrafast all-optical
modal control. Switching of polarization can be most easily achieved
with type-1 structures as in Fig. 2(a), when the optical excitation
energy is above the absorption edge of the cladding layer, but below
the absorption edge of the substrate layer. In this case, the
optical pumping will result in a substantial change of the asymmetry
factors for the two modes.

As an example, we consider an asymmetric InGaAsP waveguide
\cite{Huang} operating at $\lambda$=1.55 $\mu$m  with the core layer
index $n$=3.55, the substrate layer index $n_s$=3.24 and the UAP
cladding layer index $n_c$=3.45. Let the energy of the pump
excitation be above the cladding bandgap of $\lambda$=1.35 $\mu$m.
The resulting variation of waveguiding can be described by taking
the dielectric function of the absorbing core and cladding layers
with the Drude contribution of free carriers, viz. $\epsilon_f =
\epsilon_{f,\infty} -{\omega_{p,f}^2 / \omega^2}$ and
$\epsilon_{out_c} = \epsilon_{c,\infty} - {\omega_{p,c}^2 /
\omega^2}$, where $\omega_{p,i}^2 =N_ie^2/\epsilon_0m_{i}^*$ with
$\omega_{p,i}$, $N_i$, and $m_{i}^*$ being, respectively, the plasma
frequency, the electron-hole pair density, and the  reduced
effective mass in the core ($i=f$) and the cladding ($i=c$) layers.
The bulk optical dielectric constants of the core and cladding
materials are denoted, respectively, by $\epsilon_{f,\infty}$ and
$\epsilon_{c,\infty}$.

Linear decrease of the dielectric function of the core and cladding
layers with the free carrier concentration shifts the waveguiding
properties and the confinement factors for the two modes. The
variation of the cutoff thicknesses is shown in Fig. 6 (case 1). The
TM and TE lines intersect at $k_0d $=0.33. This means that if we
choose the core thickness $d$ to be precisely $d $=0.33$/k_0$, we
shall have only one mode confined for any pumping level. At the
pumping corresponding to $N \ge N_{c,1} \approx 8.5\times
10^{18}~cm^{-3}$ the device mode will switch from TM to TE. This
effect can be used for both polarization switching and modulation.

The switching concentration $N_{c,1}$ is sensitive to the layer
indices and can be adjusted to lower values. For the purpose of
low-power switching, more favorable structure is type-3, with both
cladding and core layers patterned as in Fig. 2(c). The cutoff
thicknesses for this case are also displayed in Fig 6 (case 2).
Making both the core and the cladding a UAP layer makes the
structure more anisotropic and the free carrier effect on the wave
propagation becomes sharper. This lowers the switching concentration
$N_{c,2}$.

It is worth noting that if one patterns only the core layer [index
profile as in Fig. 2(b)], the photoinduced free carrier
concentration would be insufficient to effect a cutoff-controlled
switch between the TE and TM modes. The variation of the asymmetry
factors is simply not strong enough in this case.

\begin{figure}
\leavevmode \centering{
\includegraphics[width=2.7in]{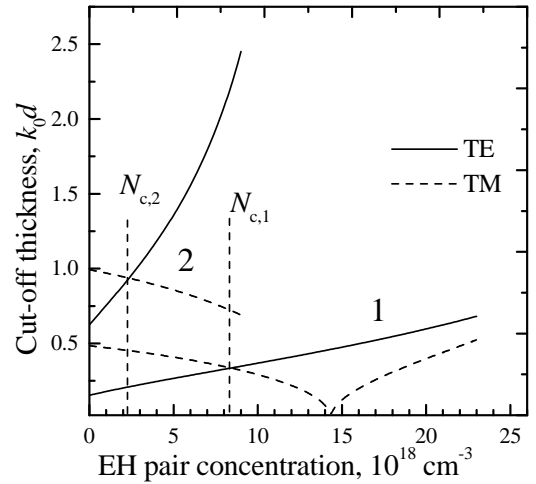}}
\caption{Cutoff thicknesses for TE and TM modes in an InGaAsP 1.55
$\mu$m waveguide structure as functions of the concentration of
optically pumped electron-hole pairs. Case (1): UAP cladding layer
only; Case (2): both the cladding and the core layers patterned. }
\end{figure}

\section{Modal control in leaky waveguides and directional couplers}
New useful polarization-dependent effects can be obtained when an
additional high-refractive-index layer is added onto the cladding
layer, or when the three-layer waveguide is placed on a
base-substrate layer of high refractive index. These effects have
been exploited in the so-called resonant-layer devices
\cite{Hammer}, vertical directional couplers and filters
\cite{Ebeling} and leaky waveguides \cite{Torner}. Uniaxial
patterning of one or more waveguide layers can provide a useful
addition to the modal control possibilities of these devices.
Here we briefly outline these possibilities. %

For waveguides on a high-index base substrate, the main effect of
the base substrate results from the exponential decay of the guided
modes due to their leakage through the bottom cladding layer into
the substrate. {\it (To avoid confusion with the already used term
"substrate" for a part of the original waveguide, we are now
referring to that layer as "bottom-cladding".}) This leakage has an
exponentially strong dependence on the difference between the modal
effective index and the index of the bottom-cladding layer, which
determines the barrier height for photon tunneling decay into the
base substrate. In a standard leaky waveguide the TE mode has a
higher index and therefore exponentially lower damping. As follows
from our discussion in the previous sections, incorporating a UAP
layer in the structure allows us to alter the bottom-cladding layer
modal transparency. This gives a variable selectivity of the
leakage-based modal control.

Adding a high-index resonant layer on the top-cladding layer with
its thickness chosen to support a mode with the same propagation
constant as the basic waveguide, leads to a well-known oscillatory
energy exchange between the two waveguides. The resonant coupling
effect underlying this exchange is exponentially sensitive to the
matching of the propagation constants. Incorporation of a UAP layer
as a core or a cladding layer, combined with the optical pumping,
enables a variable-mode vertical directional coupler that effects
fast mode selection at the time of operation.

\section{Lateral variation of UAP layer parameters}
Consider the effect of gradual variation in the density of pores in
a UAP cladding layer (vary $f$ laterally). In our limit of $\lambda
\gg a$ (even relaxed to $\lambda > a$) the effect is evidently
similar to that of lateral index variation in the cladding. It can
be used for shaping the mode field in the laser stripe, to achieve
desirable properties, similar to those obtained by the parabolic
etching of the stripe or the parabolic variation of the material
index. An example of such properties is the one-mode high-power
generation in a shaped unstable resonator laser design
\cite{Dente,Chan}. It is known that one way of obtaining a large
gain difference between the fundamental mode and higher-order modes
is to use structure profiles with a strong real-index antiguiding
and weak imaginary-index guiding. Structures with UAP layers can
provide a very effective index antiguiding. In waveguides with a UAP
core one must design the pore density so that it is highest at the
center line. On the other hand, in waveguides with a UAP cladding
layer, the antiguiding effect is achieved when the density of pores
(and hence the index contrast) grows with the distance from the
center.

We remark that while UAP layers with lateral variation $f$ offer an
effective tool for achieving high-power single-mode operation, this
approach is rather suitable only for longer wavelength, e.g., for
far-infrared devices. One needs room for smooth but sizable pore
density variation while still staying in the limit $\lambda > a$.

\section*{Conclusions}

We have derived an efficient approach to calculate the cutoff
thicknesses and optical confinement factors for a 3-layer
semiconductor optical amplifier waveguides with anisotropically
patterned layers, in which uniaxial anisotropy is deliberately
introduced in one or more of the waveguide layers.

We demonstrate that the patterned layer anisotropy can be
efficiently controlled to provide a modal control of various useful
waveguide devices. Although no attempt was made to fully optimize
the proposed devices, we show that their implementation is within a
reasonable range of lithographic and material parameters.

Finally, we note that the theoretical approach used in this work,
based on the effective media approximation in the spirit of the
Maxwell Garnett theory, has a wider range of validity than that we
have exploited so far. Thus, our approach of Sect. V can be used to
treat 2-d photonic crystals with laterally varying parameters. The
scale of lateral variation does not have to be smooth, so long as
the spatial scale of the obtained field variation in the structure
exceeds the structure pitch. For example, the same approach can be
applied to a 2-d PC with an omitted row of pores that could be
useful in the implementation of optical routers and splitters.


\appendices
\section{Guiding effects in a thin waveguide with active (amplifying or
absorbing) layers} \label{Appendix1}

Wave propagation in a 3-layer waveguide with active layers has
special features: the waves are inhomogeneous in all layers and
hence the wave propagation direction is not the same as the local
energy propagation direction. The waveguiding is described by a
system of equations, the guiding and the gain/damping effects being
interconnected.

We discuss these effects for the case of the TE mode, where we can
use the eigenvalue equation (\ref{TE-Deq}) for the propagation wave
vector in an isotropic layered structure. The TM mode can be
considered similarly.

 Consider the case of active layers, when
$\epsilon_f=\epsilon_f' + i \epsilon_f''$. Equation (\ref{TE-Deq})
and the equations for $k_{o,f},~\alpha^{c}$, and $\alpha^{s}$ now
have both a real and an imaginary part and split into pairs, e.g.:
\begin{equation}
\begin{array}{c}
 k'_{o,f}={\rm Re}\sqrt{
(\epsilon'_{\perp,f}+i\epsilon''_{\perp,f})k_0^2 - (Q'+iQ'')^2},\\
k''_{o,f}={\rm Im} \sqrt{
(\epsilon'_{\perp,f}+i\epsilon''_{\perp,f})k_0^2 - (Q'+iQ'')^2}
\end{array}
\end{equation}
The sign of the imaginary parts of the waves should be taken in the
usual manner, so that the waves go only out of the layers with
larger gain (or into the layers with larger absorption).

Thus, there are two independent variables $Q',Q''$ (the complex
propagation constant of the wave) and a system of two equations
generated by Eq. (\ref{TE-Deq}) to find them. The important point is
that {\em all} waves become inhomogeneous if at least one layer is
active, $\epsilon''_i \ne 0$, $i=s,c,f$ (excluding the case
$\epsilon''_s = \epsilon''_f = \epsilon''_c$).

All qualitative features of the waveguiding can be understood in an
exemplary case of a thin core layer, $k_{o,f} d \ll 1$, when Eq.
(\ref{TE-Deq}) becomes algebraic, viz.
\begin{equation}
k_{o,f}^2 d = \alpha^{c}+\alpha^{s}  \label{app1}
\end{equation}
Let $Q^2=k_0^2z$, $z=x+iy$. Then Eq. (\ref{app1}) reads
\begin{equation}
k_{0}d(\epsilon_f'+i\epsilon_f''-x-iy) =
\sqrt{x+iy-\epsilon_c}+\sqrt{x+iy-\epsilon_s} \label{app2}
\end{equation}

Let us first discuss the simplest case of a symmetric waveguide,
with $\epsilon_c=\epsilon_s$. For this case, equation (\ref{app2})
can be solved by introducing a dimensionless variable,
$t=\alpha^{c}/k_0=\sqrt{z-\epsilon_c}$, for which we have
\begin{equation}
k_{0}d[\delta\epsilon'+i\delta\epsilon''-t'^2-2it't''+t''^2] =
2(t'+it'') \label{app3}
\end{equation}
where $\delta\epsilon = \epsilon_f-\epsilon_c$. Then, we have
\begin{equation}
t'+ t'^2k_{0}d/2=k_{0}d(\delta\epsilon' +t''^2)/2 ,~~~t''
={k_{0}d\delta\epsilon''\over {2(1+t'k_{0}d)}}
 \label{ImRe-eq}
\end{equation}
Several important conclusions follow from Eq. (\ref{ImRe-eq}):

First,  $t'' \propto \delta\epsilon''$. Imaginary part of $t$ is,
essentially, the oscillatory contribution to the exponential decay
of the wave outside of the core. So, as is physically obvious, it is
proportional to $\delta\epsilon''$, that is to the difference
between the net gain (or loss) in the core and the cladding. This
conclusion is not peculiar to the thin layer approximation. It can
be anticipated already from the initial equation, but only for a
symmetric waveguide. For an asymmetric waveguide, the gain-loss
balance is more complicated, see below.

Second, the guiding properties of a thin-core waveguide (they depend
only on the sign and the value of $t'$) are determined by two
parameters $k_{0}d\delta\epsilon'$ and $k_{0}d\delta\epsilon''$, of
which the former defines the usual "effective one-dimensional
potential well" and the latter describes the gain (or loss) guiding
effects.

Third, we note that the guiding effects of gain or loss are
proportional to the square of the gain differences in the core and
the cladding, so that layers with step-like absorption are equally
good for guiding. However, for a thin core with gain but no index
step, the guiding [described by ${\rm Re} (t)$] is weak, being
proportional to third power of the small parameter $k_0d$.

Fourth, we note that even in the case of pure gain variation (i.e.
$\delta\epsilon' = 0$), the effect is true guiding. Indeed, the core
loses its energy only to support the growth of the wave in the
adjacent cladding layer. Inasmuch as the wave amplitude and its
energy both decrease exponentially away from the core, the guiding
can be characterized as confinement (hence we retain the possibility
to evaluate energy out of the waveguide). Interestingly, the same is
true for a lossy core, when the profile of the wave is essentially
maintained by extracting the energy from adjacent layers.

Finally, for our case of a thin core, the equations can be readily
solved perturbatively, that is by keeping in first approximation
only the lowest terms in $k_0d$.  For $\alpha_c$ this yields
\begin{equation}
\alpha_c'=k_{0}^2d\delta\epsilon'/2
,~~~\alpha_c''=k^2_{0}d\delta\epsilon''/2
 \label{GiudeGain}
\end{equation}
Therefore, in this approximation there is no gain guiding, but it
appears when higher terms in $k_0d$ are taken into account. To bring
the discussion in closer correspondence to the results of section
\ref{TE}, we note that $t$ is identical to the $\delta_{TE}$, see
Eq. (\ref{ind_TE}). We can use Eq. (\ref{ind_TE}) to write down the
modal index directly:
\begin{equation}
n_{eff,TE}=\sqrt{\epsilon_c+t^2} \approx \sqrt{\epsilon_c'} +
{{\delta^2_{TE}+i[\epsilon''_c +(k_{0}d)^2\delta \epsilon'\delta
\epsilon''/2]} \over {2 \sqrt{\epsilon_c'}}}
 \label{Index}
\end{equation}

In connection with Eq. (\ref{Index}) we make two observations: (i)
the second term in the numerator of  Eq. (\ref{Index}) gives the
well-known Dumke result \cite{Dumke} for the confinement factor (the
energy confinement has an additional factor of 2); and (ii) in a
waveguide with a thin core and weak confinement, the damping in the
cladding is more effective than the gain in the core.

For a symmetric waveguide of arbitrary core thickness, the
gain-guiding effects were recently considered by Siegman
\cite{Siegman}. Our analytical treatment is restricted to the
thin-core limit but it allows to consider the asymmetric waveguide
in a similar fashion.

For the asymmetric case we retain the definition of the cutoff as
the thickness that borders the region where the real part ${\rm Re}
(\alpha^s)$ of Eq. (\ref{app1}) vanishes. For a thin core this
cutoff thickness can be found directly from the Eq. (\ref{ind_TE}),
having in mind that at the cutoff ${\rm Re}(\delta_{TE})=0$. Damping
or gain effects shift the cutoff thickness. This results from the
quadratic in $\delta \epsilon'' $ antiguiding contribution to
$\delta_{TE}$.

An important additional issue in an asymmetric waveguide results
from the different sign of the propagative contribution to the ${\rm
Im} (\delta)$ from the asymmetry factor. Indeed, from Eq.
(\ref{app3}) we have
\begin{equation}
t = k_0d(\epsilon_f-\epsilon_s-t^2)-
\sqrt{\epsilon_s-\epsilon_c-t^2} \label{Delta}
\end{equation}
For $k_0d\ll 1$ we have $|t^2| \ll |\epsilon_f - \epsilon_s|$, so
that to the lowest order in $k_0d$ instead of Eq. (\ref{ind_TE})
above equation (\ref{Delta}) gives
\begin{equation}
\alpha_c=k_0t~~~ t = k_0d(\epsilon_f-\epsilon_s)-
\sqrt{\epsilon_s-\epsilon_c} \label{Delta1} \label{app3}
\end{equation}
Two remarks are in order here. Firstly, we observe that Eq.
(\ref{Delta1}) coincides with the reduced Eq. (\ref{ind_TE}).
Indeed, for $t$ to be $\ll 1$ the numerator in Eq. (\ref{ind_TE})
should be much smaller than the denominator, so that one can regard
the numerator as the difference of two squared and nearly equal
terms. Secondly, we note that we obtain the cutoff thickness for a
thin layer with low confinement by setting ${\rm Re}~( t)=0$ in Eq.
(\ref{Delta1}).

Separating imaginary part of Eq. (\ref{Delta1}) we obtain
\begin{equation}
{\rm Im}(\delta_{TE})=k_0d~{\rm Im}(\epsilon_f-\epsilon_s)- {\rm
Im}(\sqrt{\epsilon_s-\epsilon_c}) \label{ImDelta}
\end{equation}
For a confined mode the second term in Eq. (\ref{ImDelta}) is
smaller than the first term, but it depends on the difference
$(\epsilon_s-\epsilon_c)$ and shows the redistribution of loss
between the substrate and the cladding. For a weak confinement, the
penetration depth into the substrate is larger than that into the
cladding, so that the substrate contribution prevails. As we
decrease $d$, the first term becomes eventually smaller than the
second and the confinement is lost.

\section{Experiment of Cloutier and Xu \cite{Xu}} \label{Appendix}

These authors employed a thin 65-nm UAP  Si core layer with a
fill-factor of $f = 0.18$, a thick SiO$_2$ bottom cladding on a Si
substrate, and air for the top cladding. Under optical pumping they
observed light emission at $\lambda =1278$ nm with many
characteristics of laser radiation, unpolarized below threshold and
predominantly TM-polarized above the threshold.

\begin{figure}[h]
\leavevmode \centering{
\includegraphics[width=2.7in]{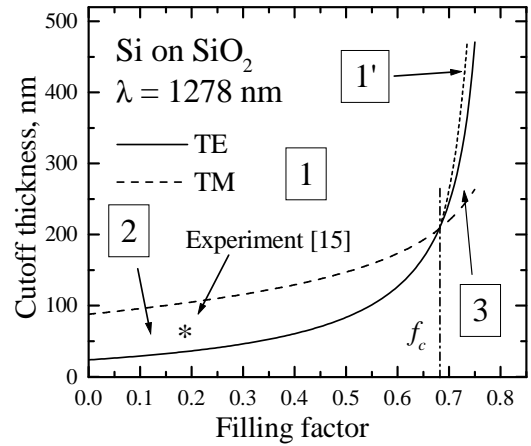}}
\caption{Cutoff thicknesses for TE and TM modes for a patterned 65
nm-thick Si layer on SiO$_2$ substrate layer as a function of
filling factor. Regions of mode confinement are denoted as in Fig.
3.}
\end{figure}

The cutoff thicknesses for the two modes at this wavelength are
displayed in Fig. 7. We see that except at very high filling factors
the TE mode is better confined. That by itself could be, in
principle, overwhelmed by other factors. For example, one could have
a peculiar radiative mechanism that favors $p$ polarization. This,
however, would also lead to TM polarization below threshold,
contrary to observations \cite{Xu}. A more reasonable explanation,
therefore, would be associated with mode-selective feedback
mechanism, e.g. higher mirror reflectivity for the TM mode.

However, the problem in interpreting the experiment \cite{Xu} is not
that the TM mode loses the competition with TE, but the fact that at
the reported parameters of the structure, the TM mode is not
confined at all. The reason for this lack of confinement is the
strong asymmetry of the air-clad waveguide.

In principle again, the absence of confinement in a passive
waveguide does not preclude the possibility that the TM mode may
still be supported by an active finite-length waveguide with gain
and feedback. In order for this to happen, the material gain should
substantially exceed the rate of leakage [in cm$^{-1}$] of the
unconfined mode into the substrate. For the TM mode this leakage
rate at the parameters of the experiment is very high (estimated to
be about $\sqrt{\epsilon_s}/d \approx 2\times 10^5$ cm$^{-1}$) and
the required gain appears quite unrealistic.

The situation would be rather different if the waveguide were made
more symmetric by adding a top cladding layer of refractive index
similar to that of the SiO$_2$ bottom cladding layer. In this case
the value of $f_c$ would become much smaller and the cutoff
thickness for the TM mode would be strongly reduced. This could be
enough to shift the structure parameters to regions $1^{\prime}$ or
even 3,
where the TM mode is dominant. %


\begin{thebibliography}{99}
\bibitem{average} We note that in the case considered
($\epsilon_{in} << \epsilon_{out} $) the "average" index $<n>$ is
rather close to $n_{\perp}$. However, in the opposite case of
$\epsilon_{in} >> \epsilon_{out} $ (dielectric cylinders in air),
the value of $<n>$ is intermediate between the true refractive
indices for the two polarizations and differs strongly from both.
\bibitem{reflection} In addition, the difference in the reflection
coefficients for the two modes also favors the TE mode generation.
It is however, less important for a long enough cavity and can be
easily taken into account.
\end{thebibliography}
\end{document}